# Hierarchical Pancaking: Why the Zel'dovich Approximation Describes Coherent Large-Scale Structure in N-Body Simulations of Gravitational Clustering

Jennifer L. Pauls and Adrian L. Melott
*Dept. of Physics and Astronomy, University of Kansas, Lawrence, KS 66045 USA*

**ABSTRACT**

To explain the rich structure of voids, clusters, sheets, and filaments apparent in the Universe, we present evidence for the convergence of the two classic approaches to gravitational clustering, the "pancake" and "hierarchical" pictures. We compare these two models by looking at agreement between individual structures – the "pancakes" which are characteristic of the Zel'dovich Approximation (ZA) and also appear in hierarchical N-body simulations. We find that we can predict the orientation and position of N-body simulation objects rather well, with decreasing accuracy for increasing large-$k$ (small scale) power in the initial conditions. We examined an N-body simulation with initial power spectrum $P(k) \propto k^3$, and found that a modified version of ZA based on the smoothed initial potential worked well in this extreme hierarchical case, implying that even here very low-amplitude long waves dominate over local clumps (although we can see the beginning of the breakdown expected for $k^4$). In this case the correlation length of the initial potential is extremely small initially, but grows considerably as the simulation evolves. We show that the nonlinear gravitational potential strongly resembles the smoothed initial potential. This explains why ZA with smoothed initial conditions reproduces large-scale structure so well, and probably why our Universe has a coherent large-scale structure.

# 1  INTRODUCTION

There can be no question that there are large-scale anisotropies in the Universe. A long-standing debate has centered on the source of this structure. When the CfA slice was first released, de Lapparent, Geller, & Huchra (1986) suggested that the observed thin sheets and large voids posed serious challenges for existing models of large-scale structure formation. Suggested explanations of this data have included cosmic strings (Stebbins, et al. 1987) and explosions that formed bubble-like structures (Ostriker & Strassler 1989).

In contrast to these somewhat exotic ideas, the Truncated Zel'dovich Approximation for gravitational clustering (hereafter TZA) offers a much simpler explanation of the structure. The Zel'dovich Approximation (ZA), calculated in co-moving coordinates, moves particles with constant velocities. Truncation of the power spectrum removes small-scale power from the initial conditions, allowing the larger modes to dominate. Gravity has the effect of amplifying any initial anisotropies. Beginning with Melott et al. (1983), such anisotropic structures were also found in N-body models of hierarchical gravitational clustering. The TZA correctly models this effect and produces the characteristic Zel'dovich pancakes. While originally the ZA was thought to be applicable only to top-down (or "pancake") models of structure formation, Coles, Melott, & Shandarin (1993) found that when the power spectrum is appropriately truncated at large $k$ before applying the approximation it works for hierarchical models as well. Melott, Pellman, & Shandarin (1994, hereafter MPS) improved the smoothing methods for the TZA by using a Gaussian cutoff in $k$-space. Melott, Buchert, & Weiss (1994) have shown that application of second-order perturbation theory results in further improvement. While the TZA models collisionless dark matter, at this level of the approximation we assume that structure in the dark matter will correspond to similar structure in the visible matter.

In this paper, we look at the later generation pancakes that form in N-body simulations with $128^3$ particles on a $128^3$ co-moving mesh. It is customary to specify power law fluctuation spectra by an index $n$ where $P(k) \equiv |\delta|^2 \propto k^n$. We had (initially) Gaussian fluctuations in the density field and $n = -3, -1, +1, +3$ power spectra evolved to $k_{nl} = 8k_f$, where $k_f$ is the fundamental mode of the box and $k_{nl}$ is the non-linear wavenumber. This yields a final effective box size of $168h^{-1}Mpc$ and particle mass of $6.3 * 10^{11} h^{-1} M_\odot$ for the $n = -1$ case and a $140h^{-1}Mpc$ box and $3.6 * 10^{11}h^{-1}M_\odot$ particle mass for the $n = +1$ case, based on scaling from the correlation length as described by Melott & Shandarin (1993). We see in Figure 1b that the structure apparent in a slice through a simulation with $n = -1$ is similar to that found in the CfA slice (de Lapparent et al. 1986) and that this structure is understandable in terms of Zel'dovich pancakes. For the purpose of this discussion, we do not distinguish between sheets and filaments, both of which form in simulations (Babul & Starkman 1992) and in the TZA.

## 2 THE ZEL'DOVICH APPROXIMATION

The Zel'dovich (1970) Approximation is:

$$\mathbf{r}(\mathbf{q}, t) = a(t)\mathbf{q} + b(t)\nabla S(\mathbf{q}). \tag{1}$$

Where $\mathbf{q}$ is the particle's initial (Lagrangian) position, $a(t)$ is the cosmological scale factor, $b(t)$ is a known function for given $\Omega$, $S(\mathbf{q})$ is the initial peculiar velocity potential, and $\mathbf{r}$ is the particle's position at time t. For $\Omega = 1$, $S(\mathbf{q})$ can be related to the initial peculiar gravitational potential, $\phi$:

$$\phi(\mathbf{q}) = -\frac{3}{2}H^2 a^3 S(\mathbf{q}). \tag{2}$$

The source term for $\phi$ in the Poisson equation is $\delta = \frac{\rho - \bar{\rho}}{\bar{\rho}}$, rather than $\rho$, to prevent divergence (Peebles 1980). We discuss the approximation in co-moving coordinates (again with $\Omega = 1$, which leads to $b = a^2$):

$$\mathbf{x}(\mathbf{q}, t) = \mathbf{q} + a(t)\nabla S(\mathbf{q}). \tag{3}$$

where $\mathbf{x}$ is the particle's co-moving position at time $t$. The deformation tensor can be defined at each particle:

$$d_{ik} = \frac{\partial x_i}{\partial q_k} = \delta_{ik} + a(t)\frac{\partial(\nabla S(\mathbf{q}))_i}{\partial q_k}. \tag{4}$$

The eigenvectors of this tensor give us the principle directions of collapse (or expansion) about a particle and the corresponding eigenvalues imply the time at which there will be infinite compression along that axis (or, if the eigenvalue is negative, that there is expansion along that axis). From the deformation tensor we also get the density from the Jacobian near a particle:

$$\rho = \frac{\bar{\rho}}{(1 - a\lambda_1)(1 - a\lambda_2)(1 - a\lambda_3)}, \tag{5}$$

where $\lambda_i$ are the eigenvalues of the deformation tensor.

The ZA as it stands is only valid for power spectra $P(k) \propto k^n$ where $n \leq -3$ (Peebles 1980, 1993a,b). The reason for this strong restriction is that as power is allowed on smaller scales, smaller objects form very early. In the ZA, non-linearity is equivalent to particles crossing the paths of other particles. Once this shell-crossing has occured, the approximation has formally broken down, since there are no forces present to slow the particles and allow them to form clumps. This breakdown occurs extremely early if $n > -3$ unless the small-scale power is removed; a reasonably late- time realization of the ZA with such a power spectrum shows no readily recognizable structures. An easy way to increase the range of spectral indices (as well as the length of time) for which the approximation is useful is to truncate the power spectrum at a wavenumber close to that going nonlinear at the time of interest. This effectively removes the small scale modes and allows the effect of the larger ones to be seen. The large modes apparently dominate in determining the position of mass; small ones are effectively virialized and not correlated with their initial values. We seek to understand this better in our paper.

The smoothing method we used for the TZA was the optimized $k$-space truncation of the power spectrum of MPS:

$$P(k) \propto k^n e^{-k^2/2k_G^2}, \qquad (6)$$

where $n$ is the spectral index, $k$ is the wavenumber, $k_G = ck_{nl}$ is the optimal Gaussian cutoff (c is indeterminably large for $n = -3$, so no cutoff is used; $c = 1.5$ for $n = -1$; $c = 1.0$ for $n = +1$; and we herein determined $c = 0.8$ for $n = +3$), and $k_{nl}$ is the non-linear wavenumber at the time of interest defined by the equation:

$$a^2(t) \int_0^{k_{nl}} P(k) \mathrm{d}^3\mathbf{k} \equiv 1. \qquad (7)$$

This removes most of the strongly non-linear behavior and allows the familiar Zel'dovich pancakes to be seen. As we shall see, these are related to the real structure that forms in N-body simulations.

## 3   COMPARISON OF THE TZA WITH N-BODY "SUPERPANCAKES"

When comparing a slice through an N-body simulation to a slice through the TZA realization of the same initial conditions, one can immediately see that large structures form in about the same places and seem to have roughly the same orientations (see Figure 1). The structures in the N-body case are known as second-generation pancakes (Peebles 1993a) or superpancakes (Kofman 1991) because in the N-body case, power is allowed on all resolved scales and small objects have already become non-linear. The larger structures are a later "generation" of structures to go through collapse.

In the TZA, structure is formed (to first order) by spheres deforming into ellipsoids. So to do a comparison to N-body models, we need a method to select spheres characteristic of the filaments. Two related methods suggest themselves. One approach is to pick a point at the center of a perceived filament and build a sphere around it. Specifically, we choose the particle nearest the center of mass of the filament in its initial (Lagrangian) coordinates. One could also look for a peak in the TZA eigenvalue field that is close to the center of mass. These peaks are "birth points" for pancakes – a natural starting point from the point of view of the TZA. Presumably these two methods will yield points separated by only one or two grid cells, and those differences could be attributed to small errors in visually choosing the filament. We tried both methods and found that while the two points *usually* are quite close and move towards each other as the simulation evolves, they do not always do so. Sometimes the TZA peak seems to be associated with a rather large clump (one of several in the filament), rather than with the filament as a whole, and the centers of mass move apart. Therefore one may not blindly decide to characterize filaments by their nearest eigenvalue field peak (although in most cases such a characterization is accurate), and we have decided to choose our spheres using the Lagrangian center of mass process.

We note that our procedure (selecting filaments rather than sheets) has introduced a bias towards objects with two axes collapsing near the same time. This choice has no special meaning beyond the fact that filaments have a higher density contrast and are therefore easier to see (Arnold, Shandarin, & Zel'dovich 1982). We will follow general verbal usage and refer to "pancaking" as the formation of anisotropic structures. Whether they are prolate or oblate depends on the ratios of eigenvalues, and the stage of evolution (e.g. see Babul & Starkman 1992).

Our procedure is fairly simple. We visually pick out a (straight) filament, and approximate it as a cylinder. We then find the particle nearest the Lagrangian center of mass of those particles which will end up in the filament. We tag the particles in a sphere around our center of mass, making sure that the radius is large enough to fairly approximate a sphere but not so large that it includes particles from outside the filament. We then follow these particles through their N-body evolution and compare their deformation to the deformation predicted by the TZA. This is illustrated in Figures 2, 3, 4, and 5 for the various spectral indices. We can see that pancaking occurs in each of these figures. The difference lies in the relative sizes of the components making up the pancakes, not in their overall shapes. We conclude that the generic pancake form occurs for each of these spectral indices.

We begin building statistics for our N-body distribution by computing the tensor (van Haarlem & van de Weygaert 1993):

$$I_{ij} = \frac{1}{n} \sum_{k=1}^{n} \left( x_{ik} - \overline{x_i} \right) \left( x_{jk} - \overline{x_j} \right). \qquad (8)$$

This tensor has the same eigenvectors as the inertia tensor, but its eigenvalues are the mean squared displacements of the points from each plane defined by a pair of principal axes. These principal axes are then compared with the eigenvectors of the TZA deformation tensor (of the particle at the center of the Lagrangian sphere). This tests whether the TZA is doing a good job of predicting the filament orientation. The ratios of the square roots of the eigenvalues from $I$ are calculated to give flattening data. For a uniform ellipsoid, it can be shown that these ratios are equivalent to the ratios of the lengths of the semi-axes.

We examined the axis pairings the eigenvalues predicted. The TZA has formally broken down for more than half of the structures examined and might be thought to be of no use in predicting these pairings. However, since the TZA eigenvalues give the relative time of collapse, their *order* ought to still be correct. Although the TZA pancakes are becoming "puffy" after shell-crossing has occured, the N-body structures are still collapsing. It is therefore reasonable to expect that the direction that the TZA gives as the earliest to collapse may be the same as the most completely collapsed direction in the N-body case at any time, even after the TZA pancake has shell-crossed. The largest compression from the TZA is matched with the shortest semi-axis of $I$.

There are a couple of ways to quantify this agreement. The most stringent test is to require all three pairs of axes to be correctly matched up (longest TZA-predicted axis with longest N-body axis, middle TZA-predicted with middle N-body, and shortest TZA-predicted with shortest N-body). Using this criterion, we find agreement in 84% of the filaments for $n = -3$; 100% of the filaments for $n = -1$; 84% of the filaments for $n = +1$; and 53% of the filaments for $n = +3$. In the $n = -3$ case, the imperfect agreement is due to the two smaller semi-axes being nearly identical in length. One would expect 17% (1/6) agreement for randomly oriented coordinate systems.

We expect better agreement if we examine only the axes which point along the most characteristic dimension (either the longest, for a more filamentary structure, or the shortest, for a more oblate structure). That is, the two axes perpendicular to the characteristic dimension of the structure could be more difficult for $I$ to distinguish, while the TZA has them "built in". (In the worst case scenario, imagine an ellipsoid with a, b, c as its semi-axes. If b=c, then the principal axes associated with b and c have completely arbitrary orientations in a plane.) When we relax our test and require only that the most characteristic axes match, the agreement improves to 100% for $n = -3$; 100% for $n = -1$; 95% for $n = +1$; and 82% for $n = +3$. In this case, random orientations would give 33% (1/3) agreement.

We can quantify the angular disagreement by finding the axis about which a single rotation will bring one coordinate system into the other. When we do this, we find the average rotation (and the standard deviation of the mean) to be $(32 \pm 7)°$ for $n = -3$; $(17 \pm 3)°$ for $n = -1$; $(29 \pm 6)°$ for $n = +1$; and $(50 \pm 4)°$ for $n = +3$. If we consider only those cases in which there was complete agreement in axis pairings (i.e. passing the more stringent of the tests discussed above), we find a reduction of roughly 10° in the average rotation (except for the $n = -1$ case, in which all the pairings were correct to begin with). For randomly oriented coordinate systems (with freedom to change the signs of the axes), the average rotation will be 60°.

Peebles (personal communication) has suggested examining axial ratios inside a collapsed Lagrangian sphere as an objective test of "superpancakes". Our ratios can be found in Figure 6. These axial ratios are remarkably similar, further substantiating our claim that pancake structures form even in models with substantial small-scale power. Since shell-crossing has already occured, or is extremely close to occuring in almost all filaments (which is to be expected for particles at the centers of filaments), the TZA cannot be used to make a numerical prediction of these ratios.

We did an additional check of the general plausibility of the TZA by looking at the amount of mass in nonlinear objects in the N-body simulation versus the TZA. The N-body slices appear to have more particles (mass) in clumps than do the TZA slices (Figure 1). For $n = -1$, which is probably closest to the slope on the scales of objects going nonlinear today, we compared the total number of particles which have shell-crossed in the TZA to the total number of particles in the simulation which ever had a negative volume element ($\mathbf{a} \times \mathbf{b} \cdot \mathbf{c}$, where $\mathbf{a}$, $\mathbf{b}$, and $\mathbf{c}$ are the vectors from a point to the current positions of its Lagrangian neighbors along each box axis). We found 36% of the TZA particles were clumped, compared to 50% of the N-body particles. We consider this to be reasonable agreement. The discrepancy is due to the fact that the TZA misses the small clumps.

We have shown that quasilinear structure in hierarchical N-body models can be approximated very simply as gravitational amplification of initial anisotropies during collapse. This was achieved by showing that when the ZA is truncated using the method of MPS, it can accurately predict where moderately nonlinear structures (like clusters and superclusters) will form, their orientations, and their relative prolateness or oblateness. We further conclude that no exotic theories are required to describe such structure formation. Gravitational amplification of Gaussian density perturbations is entirely adequate. Up to this point, we have concerned ourselves with the question of how well the TZA works. We will next turn our attention to the more interesting question of *why* it works.

## 4   WHY DO SUPER-PANCAKES EXIST?

Recent years have seen numerous numerical tests that demonstrate the similarity between the locations of clumps in hierarchical clustering models and the tracing of pancakes found by applying the ZA to the smoothed initial conditions (MPS, and references therein).

The approximate theory of adhesion (Gurbatov, Saichev, & Shandarin 1989; Weinberg & Gunn 1990; Kofman 1991; Kofman et al. 1992) provides better power spectra and mass density distribution functions than the TZA at the expense of somewhat greater phase errors (Melott, Shandarin, & Weinberg 1994). It also offers one explanation of the previously mentioned similarity. Discussions of this approximation suggest that mass will move in a smooth flow up to $R_*$, roughly the peak-to-peak length scale associated with the initial potential, providing pancakes. After this, the clumps formed move coherently in response to the potential, producing "second-generation pancakes" or "superpancakes". This stage is supposed to last until the clustering reaches the scale of the correlation length of the initial potential, $R_\phi$. Then we supposedly move into the third regime which is pure hierarchical clustering, in which motions are significantly influenced by local clumps and the effect of the initial potential is minimal. (This third stage was traditionally assumed to be the only one in hierarchical clustering theory, which never predicted filamentary structure.) The first stage is relevant only for initial fluctuation spectra which drop off sufficiently rapidly at large $k$, or have a forced cutoff as in numerical simulations.

We will present evidence that the third stage does not exist, at least for $n \leq +3$. We have seen no sudden change (only gradual improvement) as we move from models in which $R_\phi$ is much smaller than the scale of clustering ($n = +3, +1$) to those models in which $R_\phi$ is much larger than the scale of clustering ($n = -1, -3$). We have designed a severe test of the relevance of the correlation length in the initial potential in the form of an N-body simulation with an initial spectral index $n = +3$. In this extreme case the correlation length of the initial potential $R_\phi$ is extremely small, resolution limited, since the fluctuations diverge on small scales. However, as we shall see, $R_\phi$ will grow considerably as the simulation evolves. In Figure 1d, we show dot plots of thin slices of the N-body simulation and of the TZA at $n = +3$. The moment shown is $k_{nl} = 8k_f$. While the approximation is not precise, it is nevertheless obvious that there *is* a strong tendency for the positions of condensations to agree. It is useful to examine the density correlation coefficient,

$$S_{12} = \frac{\langle \delta_1 \delta_2 \rangle}{\sigma_1 \sigma_2} \qquad (9)$$

where $\sigma_i \equiv \langle \delta_i^2 \rangle^{1/2}$ as plotted in Figure 7. The cross-correlation of the $n = +3$ N-body density field with the TZA is worse than the other spectra, as one would expect, but far better than its cross-correlation with Eulerian linear theory. This shows that "pancake" dynamics are operative to some extent on large scales for $n = +3$. The important difference here is that in our $n = +3$ simulation the correlation length of the initial potential is vastly smaller than the scale of clustering, and long-range initial correlations are of extremely small amplitude for $n = +3$. Nevertheless, they appear to dominate large-scale behavior. It has generally been assumed that coherence in motion would be nearly absent for such a case. Our results suggest that it is important to consider instead the correlation length of the evolved potential. When one does this, it is understandable how long wave modes of the initial potential are able to dominate motion on scales much larger than the initial $R_\phi$. Strong correlations exist for all $n$, increasing for decreasing $n$.

We suggest that the transition from pancake to hierarchical regime takes place gradually over the entire range of spectral indices from $n = -3$ to $n = +4$. For $n \leq -3$, collapse on all scales happens nearly simultaneously which is a noncontroversial sufficient condition for pancaking (Peebles 1980, 1993b). For $n \geq +4$, the minimal $k^4$ tail from dynamics (Zel'dovich 1965, Peebles 1980 §28) ought to dominate over initial power. In this case, any kind of perturbation theory applied to the initial conditions ought to break down, since large-scale power at late times should be almost unrelated to large-scale power in the initial conditions. Figure 7 shows the onset of this breakdown at $n = +3$; at this index, the cross-correlation does not clearly approach 1 for small $\sigma$. Previous studies (Beacom, et al. 1991; Melott & Shandarin 1993) have shown that the primary visual change with index $n$ is based on the changing mass of clumps. At $n = -3$, the pancakes are assembled from essentially unclustered matter. As the index $n$ increases, gradually larger clumps are seen. For the $n = +3$ case studied here, there are perhaps two big clumps per pancake which makes it problematic to speak of filaments. Yet the position, and to a somewhat lesser degree the orientation, of these clumps are accurately predicted by the TZA. This means there is coherent motion on scales far exceeding the correlation length of the initial potential. The extremely low-amplitude power present on very large scales is still enough to dominate the motion up to $n = +3$.

We suggest a simple explanation: initial power at high wavenumbers has been essentially "thermalized", going into binding energy inside clumps (Peebles 1980 §28). Feedback from small scales to large is extremely weak, and as we know from past numerical work the growth rate of nonlinear modes is suppressed. This acts like smoothing, making the small-scale power irrelevant to dynamics outside clumps. Of course, the mass distribution is not smooth: the high frequency power exists. But for bulk motion, what counts is power on large scales, which is accurately described up to mildly non-linear regimes by the TZA.

## 5    EVOLUTION OF THE GRAVITATIONAL POTENTIAL

There have been hints before that gravitational clustering is like smoothing (see Press & Schechter 1974; Melott, Weinberg, & Gott 1988; Weinberg & Gunn 1990; Beacom, et al. 1991; Little, Weinberg, & Park 1991; Evrard & Crone 1992; Kofman 1991; Kofman, et al. 1992). This suggests an explanation for the success of the TZA, which consists of moving particles by the ZA in the smoothed potential of the initial conditions. We will show that it is more than an analogy: the nonlinear evolution may be quite reasonably described by Gaussian smoothing; $n = +1, +3$ the smoothed initial potential looks nothing like the unsmoothed initial potential but does resemble the *unsmoothed evolved* potential. For $n = -1, -3$ the potential does not evolve much, so the smoothed and unsmoothed initial and unsmoothed final all resemble one another. This could explain why the TZA works outside the regime in which it was expected to.

In Figure 8 we show one diagonal cut through the gravitational potential of the initial and evolved ($k_{nl} = 8k_f$) simulation cubes superimposed. Recalling that the potential scales as the expansion factor to linear order, we have appropriately scaled down the evolved potential. For $n = -1, -3$ the potential evolves very little. This is because the power spectral index of the potential is $n - 4$, from the Poisson equation; it is dominated by the low-$k$ linear modes in both cases. Of course the potential we see here is dependent on the low-$k$ cutoff (the fundamental), and a larger box would immediately show a larger oscillation. This would, however, only reinforce the similarity between initial and evolved potential. This lack of evolution makes all sorts of approximation schemes work well (Melott 1994) for $n \leq -1$.

For $n = +1, +3$ the situation is quite different. The divergences in the gravitational potential are only logarithmic for $n = +1$ but are strong at high-$k$ for $n = +3$. The initial potential shows the usual high frequency oscillations and is dominated by them for $n = +3$. More resolution would give rise to even more violent oscillations. However, in this case the unsmoothed evolved N-body potentials are very different. They display almost no hint of the fine structure present in the initial conditions. In this sense gravitational clustering has "smoothed" the potential. Can we push idea this a step further? In Figure 9 we explicitly compare the *smoothed* initial potential to the *unsmoothed* evolved potential. We used Gaussian $k$-space smoothing, as we used for the TZA (see equation (6)). The agreement is remarkable, considering the deeply nonlinear evolution and the very small initial long-wave amplitude.

In order to make slightly more quantitative our claim about the potentials, we have computed the cross-correlation coefficient (see equation (9)) for the peculiar gravitational potentials on the $128^3$ mesh. In Table 1 we show the cross-correlations between pairs of the three potentials for each spectral index – initial, optimally smoothed initial, and final ($k_{nl} = 8k_f$). The case of $n = -3$ is shown for reference only. The initial potentials were smoothed using a Gaussian smoothing length which maximized their cross-correlation.

First, for $n = -3, -1$, all cross-correlations are large. This is a simple consequence of the strong emphasis on small-$k$ modes in the potential. There is a small improvement in the third significant figure for $n = -1$. But the cases of $n = +1, +3$ are more interesting. In these cases the smoothed initial potential is *much* more strongly correlated with the nonlinear potential than is the raw initial potential. This is not just an artifact of removing high-frequency power, as shown by the cross-correlation of the smoothed initial potential with itself (unsmoothed). This is a quantitative measure of the fact that the evolved potential resembles the smoothed initial potential, as seen in Figure 9. Although they are related by the Poisson equation, this agreement is not true for the mass density: the evolved mass density does *not* resemble the linearly amplified smoothed initial mass distribution (Coles, et al. 1993). Those correlation coefficients were less than 0.1 for $n = -1, +1$. Beyond the improvement one sees in the cross-correlation with smoothing, there is apparent visual improvement from Figure 8 to Figure 9. In Figure 8 for $n = +1, +3$, one sees many more oscillations in the initial potential than in the final. However, after smoothing the initial potentials (Figure 9), the initial and final potentials show very nearly the same number of peaks and valleys.

The ZA uses the potential to move the particles. The fact the smoothed initial potential resembles the evolved potential even for $1 \leq n \leq 3$ helps explain the unexpected accuracy of the TZA. The potential in which the clumps move has a lot of long-range coherence. This becomes true, even if it wasn't true in the initial conditions. Therefore, one need not worry about the extremely small initial coherence length for $n = +1, +3$ – by the time clumps have formed, the potential has greatly increased its coherence length and *that* is the potential in which the clumps move.

In Table 2 we show the coherence length of the potential in the initial conditions and at the stage $k_{nl} = 8k_f$. We define it as the radius at which the auto-correlation falls to half its value at the smallest radius we can resolve. For $n = +3$, the coherence length is entirely dependent on the large-$k$ cutoff, and will become arbitrarily small as the resolution is increased. For $n = -1, -3$ the coherence length will become large as the box size is increased, since in these models it is dominated by the long waves.

It is changes that we seek. For $n = -3, -1$, the coherence length is unchanged. In both cases the numerical limit is at small $k$. For $n = +1, +3$ the coherence length increases by a factor of 7 to 8 as gravitational clustering proceeds. The initial values are quite small and the evolved values are nearly equal to the lengths of the filaments which were visually selected. This coherence of the evolved potential shows why bulk motions are coherent and lead to "superpancakes".

We have shown smoothed potentials and cross-correlations for the potential-optimized smoothing lengths. In principle, the potential-optimizing wavenumber, $k_\phi$, is independent of $k_G$, which optimizes the density field agreement upon implementing the TZA, as found by MPS. But, in fact, these wavenumbers are related. For $n = -1, +1$, the evolved potential in the N-body simulation agrees best with the initial potential smoothed with $k_\phi$ equal to or a bit larger than $k_{nl}$ but still smaller than $k_G$. For $n = +3$, $k_G$ and $k_\phi$ were a bit smaller than $k_{nl}$ and were the same within errors. This is because for $n = +3$ the approximation is breaking down as $\phi$ diverges more strongly ($\propto k^2$) and the Gaussian cutoff must move toward smaller wavenumbers to "contain" it. (It's only fair to point out that for $n = -1$, the correlation coefficient varied *very* little with wavenumber in the range $4k_f$ - $64k_f$.)

The TZA works because as it is used to approximate any stage, one moves the particles in a potential which is very nearly the correct evolved potential. The small-scale objects are not correctly formed, as they have been smoothed away, but the structures currently becoming non-linear *are* correct, because the potential is (nearly) correct for them. The current arrangement of clumps in the N-body simulations show pancakes from past motion, and their future will drive them into larger ones. At any stage the N-body potential is in fact smooth on the same or larger scale as the current generation of "superpancakes" is (as characterized by the scale associated with $k_G$), and therefore these objects will move coherently to make the next generation. Thus we have "hierarchical pancaking".

This result suggests new approximations which may be worthwhile. Particle-pushing versions of adhesion (Weinberg & Gunn 1990) as well as the Frozen Potential Approximation (Brainerd, Scherrer, & Villumsen 1993; Bagla & Padmanabhan 1994; Sathyaprakash et al. 1994) move particles in some way in the initial potential (or its evolved approximation). New improved versions of these approximations may be constructed which move the particles in steps in the variably *smoothed* initial potential. This is not to suggest that they will be cost-effective replacements for N-body simulations, but they should add to our understanding of clustering.

The agreement of the nonlinear potential with the smoothed initial potential is nontrivial. The evolved density field does *not* resemble the smoothed initial density field (Coles, et al. 1993) or the initial potential field, either smoothed or unsmoothed. The nonlinear part of the evolved density power spectrum has a spectral index around $n = -1$, regardless of its initial value of $n$ (e.g. Melott & Shandarin 1993). This makes the nonlinear part of the potential insignificant, since it has the $(n - 4)$ dependence. The lack of any significant propagation of information from small to large scales, when combined with this, can explain the resemblance of the potential to its smoothed antecedent. Although the result is in this sense not surprising, it nevertheless unifies diverse hints that gravitational clustering acts like smoothing.

# 6 CONCLUSIONS

The Zel'dovich Approximation based on the smoothed initial gravitational potential (TZA) works with continuously changing accuracy over the range of spectral indices $-3 \leq n \leq +3$. There are no sudden jumps in behavior, so that the transition from pancaking to hierarchical clustering is continuous. The TZA can be used to predict (via the eigenvalues and eigenvectors of the deformation tensor) the orientation and relative oblateness or prolateness of filaments that arise in N-body simulations of hierarchical clustering.

The gravitational potential in an evolved hierarchical clustering simulation closely resembles the smoothed potential of the initial conditions, even though this is not true of the mass density. This smoothed potential justifies the use of the Zel'dovich Approximation with smoothed initial conditions (TZA). A new understanding of galaxy and large-scale structure formation can be developed by combining the hierarchical and pancake approaches which were formerly thought to be mutually exclusive. By doing this, we can explain the coherence of motion which leads to hierarchical pancaking, and provide an explanation for the visually rich large-scale structure of the Universe.

## Acknowledgments

We are grateful for U.S. financial support from NSF Grant AST-9021414, NASA grant NAGW-3832, and computing at the National Center for Supercomputing Applications. It was useful to exchange ideas with all the participants at the June 1994 workshop at the Aspen (Colorado) Center for Physics. We thank Jim Peebles for encouragement and suggestions, David Weinberg for discussions of adhesion theory, and Bob Scherrer for a discussion of potentials. Thanks also to Keith Ashman and Christina Bird for helpful suggestions.

**Table 1.** Cross-correlation of gravitational potentials.

| Spectral Index | Smoothed Initial/Initial | Initial/Final | Smoothed Initial/Final |
|---|---|---|---|
| -3 | – | 0.96 | – |
| -1 | 0.999 | 0.987 | 0.990 |
| +1 | 0.68 | 0.65 | 0.94 |
| +3 | 0.14 | 0.10 | 0.69 |

**Table 2.** Coherence length (in cell units) of the gravitational potential for the spectral indices discussed.

| Spectral Index | Initial | $k_{nl} = 8k_f$ |
|---|---|---|
| -3 | 36.0 | 36.0 |
| -1 | 31.5 | 31.5 |
| +1 | 1.7 | 7.6 |
| +3 | 0.6 | 4.2 |

**Tables**

**Figure Captions**

Figure 1. Thin slices taken from N-Body (left) and TZA (right) realizations with the same (random) phases with spectral indices. The moment shown is $k_{nl} = 8k_f$; where $k_f$ is the fundamental mode of the box. Figure 1a: $n = -3$; Figure 1b: $n = -1$; Figure 1c: $n = +1$; Figure 1d: $n = +3$.

Figure 2. Projections of the particles initially in a sphere chosen by the method detailed in Section 3. The three columns show the projections along the three eigenvectors calculated using equation (7). The four rows show these projections at four stages: initial positions, $k_{nl} = 32k_f$, $k_{nl} = 16k_f$, and $k_{nl} = 8k_f$. The filaments were selected when $k_{nl} = 8k_f$. The initial spectral index is $n = -3$.

Figure 3. Same as Figure 2, but with initial spectral index $n = -1$.

Figure 4. Same as Figure 2, but with initial spectral index $n = +1$.

Figure 5. Same as Figure 2, but with initial spectral index $n = +3$.

Figure 6. Axial ratios with error bars. The filled triangles show the ratio of the longest to shortest semi-axes and the open triangles show the ratio of the middle to shortest axis. Error bars show the standard deviation.

Figure 7. Cross-correlation of the smoothed N-body simulations at $k_{nl} = 8k_f$ with the smoothed TZA against $\sigma$ of the smoothed N-body density field. In decreasing order of amplitude, $n = -3$ (long dashes), $n = -1$ (short dashes), $n = +1$ (dot-short dash), and $n = +3$ (solid). The lowest line (long dash-short dash) is a cross-correlation of the $n = +3$ simulation with a linear theory extrapolation from its initial conditions.

Figure 8. The peculiar gravitational is shown along one diagonal of each of the four simulation cubes (a) $n = -3$, (b) $n = -1$, (c) $n = +1$, (d) $n = +3$. Note that there is strong evolution in the potential for $n = +1, +3$, but not for $n = -3, -1$. Solid line shows the potential evolved to $k_{nl} = 8k_f$ and scaled according to linear theory. Dotted line shows the initial potential. The units are arbitrary.

Figure 9. The peculiar gravitational potential is plotted in the same manner as in Figure 7, except that the initial potential only has been smoothed with a Gaussian optimized for the potentials. The smoothed initial potential strongly resembles the unsmoothed potential in the nonlinear regime. Note: for the case of $n = -3$, the optimal smoothing was no smoothing, so Figure 8a is identical to Figure 7a; also, the vertical scales have been changed for Figures 8c,d as compared with Figures 7c,d.